# Low-temperature condensation of carbon

S. A. Krasnokutski[1], M. Goulart[2], E.B. Gordon[3], A. Ritsch[2], C. Jäger[1], M. Rastogi[2], W. Salvenmoser[4], Th. Henning[5], P. Scheier[2]

[1] Laboratory Astrophysics Group of the Max Planck Institute for Astronomy at the Friedrich Schiller University Jena, Institute of Solid State Physics, Helmholtzweg 3, D-07743 Jena, Germany

[2] Institute for Ion Physics and Applied Physics, University of Innsbruck, Technikerstr. 25, A-6020 Innsbruck, Austria.

[3] Institute of Problems of Chemical Physics, Russian Academy of Sciences, Chernogolovka, Moscow oblast, 142432 Russia.

[4] Institute of Zoology, University of Innsbruck, Technikerstr. 25, A-6020 Innsbruck, Austria.

[5] Max-Planck Institute for Astronomy, Königstuhl 17, 69117 Heidelberg, Germany

**5/4/2018**

**Abstract:** Two different types of experiments were performed. In the first experiment we studied the low-temperature condensation of vaporized graphite inside bulk liquid helium while in the second experiment the condensation of single carbon atoms together with $H_2$, $H_2O$, and CO molecules inside helium nanodroplets was studied. The condensation of vaporized graphite leads to the formation of partially graphitized carbon, which indicates high-temperatures, supposedly higher than 1000 °C, during the condensation. Possible underlying processes responsible for the instant rise in temperature during condensation are discussed. This suggests that such processes cause the presence of partially graphitized carbon dust formed by low-temperatures condensation in the diffuse interstellar medium. Alternatively, in the denser regions of the ISM, the condensation of carbon atoms together with the most abundant interstellar molecules ($H_2$, $H_2O$, and CO), leads to the formation of complex organic molecules (COMs) and finally organic polymers. Water molecules were found not to be involved directly in the reaction network leading to the formation of COMs. It was proposed that COMs are formed via addition of carbon atoms to H$_2$ and CO molecules (C + H$_2$ → HCH, HCH + CO → OCCH$_2$). Due to the involvement of molecular hydrogen, the formation of COMs by carbon addition reactions should be more efficient at high extinctions compared with the previously proposed reaction scheme with atomic hydrogen.

**Introduction**

A possible solution to the lifetime problem of interstellar dust is to assume its low-temperatures growth directly in the interstellar medium (ISM) (Draine 2009). The lifetime of silicate dust in the ISM is a subject of great uncertainty, while the carbonaceous dust, due to its much lower stability, has to be rapidly recycled in the ISM (Jones & Nuth 2011). The temperature of the ISM could be as low as few Kelvins (Lippok *et al.* 2016). A possibility of the silicate dust formation at low temperatures was recently demonstrated and its properties investigated (Krasnokutski *et al.* 2014; Rouillé *et al.* 2014a; Rouillé *et al.* 2014b). At the same time, not much is known about the properties of the carbon material produced by condensation at low-temperatures. Some information is available due to the studies of the formation of diamond-like carbon by the physical vapor deposition technique (Robertson 2002). In this method the carbon vapors are deposited on a substrate, which is kept at room or low temperatures. However, in order to reach a high hardness (due to sp$^3$ bonding), a high energy (~100 eV per C atom) of the gaseous species is commonly used. Such non-equilibrium conditions are not very common in astrophysical environments. Studies on the low-temperature homogeneous nucleation of carbonaceous dust or low-temperature condensation of carbon on substrates are very limited. The condensation of carbon was studied using the matrix isolation technique (Wakabayashi *et al.* 2004). Vaporized graphite together with an inert gas was deposited on a low temperature substrate. In the formed matrix the carbon molecules are initially isolated. Upon annealing and slow evaporation of the matrix gas, the trapped molecules gain mobility and start to aggregate. The evaporation of Ne matrix was followed by flashes, the optical emission spectra of which matched with the theoretical Planck curves of the blackbody radiation for the temperature of 2500 K. The authors attributed the temperature rise to the energy released due to the rearrangement of bonds from the metastable sp-carbon chains to the more stable $sp^2$ bonding material. At the same time, there are many reports that carbon materials with polyyne chains are stable (see, for an example, Casari *et al.* (2016) and references therein). It was also proposed that in the case of coagulation of small nanoparticles, the release of binding energy should be sufficient to heat the formed nanoparticles to the temperature of a few thousand Kelvins (Gordon *et al.* 2011). Recently, this conclusion was also confirmed by monitoring the light emission from metal nanoparticles grown inside superfluid helium (Gordon et al. 2017b). In the case of carbon both effects could be important. The temperature rise, due to the release of the binding energy, could be a trigger for the rearrangement of metastable sp-carbon chains. Such a temperature rise, independent of its origin, could considerably modify the properties of carbon. However, the formation of completely crystalline grains, as observed for metal nanoparticles grown at low temperatures (Gordon *et al.* 2017a), is not expected for carbon. In the case of carbon, its extremely high vapor pressure over the liquid prevents melting. However, a partial graphitization of carbonaceous nanoparticles could take place even at temperatures much below the melting point (Barbera *et al.* 2014) and therefore should be expected. It can lead to an appearance of a partially graphitized carbon as opposed to a completely amorphous material that is normally expected for low-temperature condensation. In this article, we study the properties of the solid carbon material produced by the low-temperature condensation of vaporized graphite or by reactions and condensation of single carbon atoms with molecules most abundant in interstellar environments. The condensation was studied inside liquid helium, which absorbs the reaction energies allowing associative reactions. Therefore the reactions on the surface of the cold dust grains are simulated.

**Experimental**

The condensation of carbon vapor in bulk liquid helium was performed in the experimental setup described earlier (Gordon, et al. 2011). In this experiment, the graphite target was mounted inside bulk liquid helium kept at a temperature of about 1.7 K. Light from a pulsed NdYag laser ($\lambda$ = 1062 nm, E = 0.1 mJ, $\tau$ = 0.5 ns) was focused on the target surface to a spot of ~50 µm in size. This leads to the evaporation of graphite and the dominant formation of $C_3$ molecules. Some minor quantity of $C_2$ molecules as well as single carbon atoms are also produced (Kaiser & Suits 1995; Krasnokutski & Huisken 2014a). The condensation of carbon took place inside liquid helium and the formed carbon particles precipitate on the bottom of the cryostat, where Lacey carbon transmission electron microscopy (TEM) grids were mounted. After finishing the deposition and warming up the cryostat, the grids were removed from the cryostat and the deposit was analyzed by high-resolution transmission electron microscopy. The photo of submersible experimental cell used in the experiments is available in the appendix.

The condensation of single carbon atoms was performed inside nanosized helium droplets using the experimental setup described elsewhere (Krasnokutski *et al.* 2016). The principle scheme of the experiment is given in the appendix. Helium droplets were produced by expanding compressed helium gas (99.9999% purity) thorough a cooled nozzle into a vacuum chamber. The selection of helium droplets sizes were performed by adjusting the temperature of the nozzle (Toennies & Vilesov 2004). After passing through a skimmer, the helium droplets arrive to the next differentially pumped chamber. In this chamber, the helium droplets were doped with single carbon atoms using an atomic carbon source (Krasnokutski & Huisken 2014a). Additionally, the helium droplets were doped with $H_2O$ and CO molecules from residual gas ($p$ = 3×10$^{-7}$ mbar) and with $H_2$ molecules provided through a leak valve from outside ($p = 2\times10^{-6}$ mbar). After approximately 1 ms the doped helium droplets collided with the substrate. Upon collision, the liquid helium is evaporated, leaving a refractory condensate on the substrate. Graphene coated TEM grids and fused silica substrates were used for the TEM analysis and spectroscopic measurements, respectively. The deposit was analyzed using transmission electron microscopy, which was performed using a JEOL JEM 3010 microscope with a $LaB_6$ cathode operating at an acceleration voltage of 300 kV. A Fourier transform IR spectrometer (Bruker VERTEX 80v) and a UV spectrometer (JASCO V-670 EX) were used to obtain spectral data. Additionally, due to limited transparency of the fused silica material in the IR range, the measurements were performed in a broader spectral range as follows: The substrate was scratched with a thin stainless needle; After addition of a considerable amount (~30 µm pile) of a condensate, the needle was placed under an IR absorption microscope (A590), which is an accessory of a Fourier Transform Infrared Spectrometer (IFS 113v). The needle was oriented in a way that the IR beam could pass through the attached material without interacting with the needle. We used the smallest (30 µm) aperture of the microscope. In this way, the IR spectra were obtained for a free standing pile of the condensate material. To study the chemical composition and the structure of solid deposits, an Auriga 60® CrossBeam Workstation equipped with a spectroscopy Oxford X-Max energy-dispersive X-ray (EDX) detector for 2D und 3D microanalysis was used. To minimize the interaction volume of electrons with the condensate, a relatively low accelerating voltage of 5 kV was used. Raman spectra were measured by a micro Raman spectrometer (LabRAM 1) using an excitation wavelength of 638 nm. The B3LYP hybrid functional with the 6-311+G(d,p) basis set implemented in the GAUSSIAN09 package were used to optimize the molecular geometries of reactants and products of chemical reactions and to search for local extrema in the potential energy surfaces of reaction pathways. These calculations were used to test the presence of energy barriers in the reaction pathways. More details on the computations used in the study are available in the appendix.

**Results**

   **Condensation of vaporized graphite in bulk liquid helium**

Figure 1 shows the TEM images of the carbon material produced by the condensation of vaporized graphite inside bulk superfluid liquid helium. As can be seen in Figure 1a, the carbon is arranged in the form of wires, which is due to the condensation of carbon molecules and clusters along the cores of the quantum vortices (Gordon, et al. 2011; Gomez *et al.* 2014). This results in the formation of carbon wires of about 5 nm diameter. Upon settling on the substrate, these wires may coagulate to form a larger bulk (not shown in the figure) material or stay separately. A higher resolution image (right panel) of the formed carbon wires shows partial graphitization of the material. The formation of fullerene- or onion-like structures was observed. The comparison with previous gas-phase condensation experiments (Huisken *et al.* 2009; Jäger *et al.* 2009) shows that the structure of carbon obtained in such a low-temperature condensation is rather similar to that produced at temperatures higher than 3500 K. This is supposed to be due to a rise of the temperature during the condensation. This confirms previous observations (Wakabayashi, et al. 2004, Yamaguchi & Wakabayashi 2004, Gordon, et al. 2017b). In contrast to the metal nanoparticle formation, the complete crystallization of carbon does not occur. Partial graphitization of carbonaceous nanoparticles could happen at rather low temperatures above 600 °C (Barbera, et al. 2014). Therefore, independent of the nature of this temperature rise, it can seriously affect the morphology of carbonaceous grains formed by the low-temperature condensation process, leading to the appearance of partially graphitized grains. It should be mentioned that the level of graphitization of carbon in our sample changes in the broad range. Beside material shown in Figure 1 we observed some areas of completely amorphous carbon same as the areas of carbon with much longer graphene layers.

**Condensation of atomic carbon together with common ISM molecules inside helium nanodroplets**

Most of the models dealing with the destruction of carbonaceous dust expect complete atomization of the dust leading to the formation of single carbon atoms rather than $C_3$ molecules, which are dominantly formed after evaporation of graphite (Krasnokutski & Huisken 2014a). The expected atomization of carbon is in line with a very high fraction of carbon present in the form of atomic gas, which could be up to 80% of all carbon present in the ISM (Snow & Witt 1995). At the same time, the atomic carbon is much more reactive compared to carbon molecules. Fast low-temperature reactions between carbon atoms and many carbon bearing molecules have been previously reported (Chastaing *et al.* 2001; Costes & Naulin 2010). Therefore, the low-temperature condensation of carbon atoms can give completely different results. In particular, it could be important in the case of condensation of carbon atoms together with other species on the surface of interstellar ice grains, which are expected to be present in molecular clouds.

To confirm this hypothesis, we studied reactions of single carbon atoms together with $H_2$, $H_2O$ and CO molecules inside liquid helium droplets. The nanoparticles produced in helium droplets were deposited on two different substrates. For HRTEM studies, graphene layers on Lacey carbon grids were used. As can be seen in Figure 2, almost round nanoparticles with a diameter of ~5 nm are formed inside the helium droplets. Considering the fact that these particles were deposited on the graphene layers, which were also composed of carbon atoms, an exact chemical analysis of the deposited particles was not possible. To gain information about the chemical composition of the condensate, EDX analysis of the material deposited on the fused silica substrate was performed using a field emission scanning electron microscope. Due to the non-carbonaceous substrate and a much thicker layer of the deposit, EDX was expected to yield better results. The results of this analysis are given in Table 1. The carbon present on the $SiO_2$ substrates contains a large fraction of oxygen. The amount of oxygen was estimated around 25 % of the total mass of the material. However, a small amount of detected silicon (~ 1%) is due to the $SiO_2$ substrate. Therefore, it cannot be excluded that some small fraction of the detected oxygen is also due to the substrate. The amount of hydrogen was impossible to quantify due to limitations of the EDX technique. Nitrogen is completely missing in the condensed compound. If other elements (nitrogen, sulfur, etc.) are present during condensation, their efficient incorporation into the solid condensate is expected. These measurements reveal the formation of some organic material with a large number of incorporated oxygen. To better characterize the property of this material, IR absorption and Raman spectroscopy were performed. The Raman spectroscopy measurements were unsuccessful due to an extremely high luminescence yield of the produced material. In spite of the decrease in the luminescence intensity of about one order of magnitude during the first 5 minutes of the measurements, the luminescence after five hours of irradiation of the condensate was still too strong to allow Raman measurements. The high luminescence yield and the fast decomposition of the material caused by irradiation with the Raman excitation laser is already an indication for the presence of complex organic compound in the sample. It also points to the fact that the material as it is produced by condensation is not very stable and should be strongly modified by UV irradiation present in the interstellar environments. The measured IR absorption spectra are given in Figure 3. The measurements performed with the substrate are shown in red. The free standing material spectra measurements (shown in black) were performed about one month later. As can be seen from the figure, the band at 3283 cm$^{-1}$, associated with the ≡CH stretching vibration, demonstrated the relative intensity decrease in the later spectrum. This is likely associated with the slow rearrangement of polyyne chains in the material. This can also explain missing bands of the C≡C stretch vibrations in the later spectrum. At the same time, the weak broad band of the hydrogen bonded OH groups (3430 cm$^{-1}$) shows a slight increase in the intensity, which can be due to adsorbed water molecules. The most intense bands in the spectrum are 2847 and 2914 cm$^{-1}$, which are due to the =CH$_x$ stretch vibrations, and intensity of the aromatic CH stretch band (3085 cm$^{-1}$) is very low. The presence of oxygen is demonstrated by the intense C=O stretching band at 1734 cm$^{-1}$ and several C-O bands in a range between 1000 and 1250 cm$^{-1}$. Bands in a range of 1350 – 1620 cm$^{-1}$ are due to the C=C stretching vibrations. The large number of bands in the spectra points to a variety of complex organic molecules formed by the low-temperature condensation of carbon atoms together with the H$_2$O, H$_2$ and CO molecules.

**Discussion and astrophysical implication**

The detection of the formation of COMs by the low-temperature condensation of carbon atoms together with the H$_2$, H$_2$O and CO molecules is an important finding. The COMs were detected in many astrophysical environments, and in particular, in the low-temperature parts of the ISM (Belloche *et al.* 2013; Hollis *et al.* 2000; Hollis *et al.* 2006; Requena-Torres *et al.* 2008). As many of these COMs are involved in metabolism of living bodies, their formation in space attract much attention as a possible step in the route towards the origin of life. Most commonly, it is supposed that the formation of these molecules proceeds in the solid state, either on the surface of dust particles or inside of molecular ices (Herbst & van Dishoeck 2009). However, there is a lack of complete understanding how these molecules are finally transferred into the gas phase. Therefore, the formation through a sequence of bottom-up gas-phase reactions was also proposed (Vasyunin & Herbst 2013).

Experimentally, the formation of COMs was found after irradiation of molecular ices with energetic photons or ions (Agarwal *et al.* 1985; Ioppolo *et al.* 2011; Vidali *et al.* 2004) or by simple addition reactions (Fedoseev et al. 2017; Chuang *et al.* 2017; Linnartz *et al.* 2015). Our findings demonstrate a new pathway in the formation of COMs by the low-temperature addition of carbon atoms to the $H_2$ and CO molecules. The formation of COMs by addition of C atoms has already been considered (Herbst & van Dishoeck 2009; Ruaud *et al.* 2015). Inside dark molecular clouds the $C^+ \rightarrow C \rightarrow CO$ transition takes about $10^6$ years (Phillips & Huggins 81). Therefore, on the early stages of molecular clouds, the carbon chemistry is dominated by reactions of C atoms and ions. The atomic carbon is found to relate to the dense molecular gas (Beuther *et al. 2014)*. Later, in the center of dense molecular clouds, the abundance of atomic carbon decreases to a constant level, which is maintained mainly due to dissociation of CO molecules by Lyman-alpha photons. Inspite of the decrease in the carbon atom abundancy at later stages, Ruaud *et al.* (2015) proposed that the formation of COMs via addition of carbon atoms is always the dominant pathway.

A comparison of the initial stages of COMs formation by carbon atom addition proposed previously and in this study is given in Figure 4. The carbon atom addition pathway proposed for the formation of COMs (Herbst & van Dishoeck 2009; Ruaud *et al.* 2015) requires the presence of hydrogen atoms and therefore is not expected to be active in our experiments. The hydrogen atoms cannot be picked up by liquid helium droplets and therefore their reactions can be completely excluded from the consideration. At the same time, dihydrogen is efficiently picked up by helium droplets and therefore, it is expected to participate in the reactions. An efficient inclusion of hydrogen into the formed COMs is demonstrated by the intense bands of the CH stretching vibrations observed in the absorption spectra. The reaction $C + H_2O$ has some energy barrier and encounters of these reactants at low-temperatures are expected to produce only weakly bound complexes (Schreiner & Reisenauer 2006; Krasnokutski & Huisken 2014b). Therefore, water molecules should not be directly involved in the reaction network. This is also confirmed by our measurements, as a very weak band associated with OH vibrations was observed, while the reaction with water molecules would lead to formation of a large number of OH groups. That is why the pathway suggested by Ruaud, et al. (2015), which included reactions with water molecules, is not expected to be active. At the same time, the reaction C + $H_2$ + M → HCH + M, where M is the third body, is barrierless (Krasnokutski *et al.* 2016). It can be followed by the CO + HCH + M → $OCCH_2$ + M reaction, which is predicted to be barrierless based on our quantum-chemical computations and then continues by the addition of the C, CO and HCH species. Addition of oxygen to the end of the carbon chain makes such a molecule unreactive toward the CO molecule. However, it still can react barrierlessly with C atoms. If C atoms approach from the side of $CH_2$ group, the barrierless insertion of C atoms inside the CC bond takes place. However, if a C atom approaches in the direction close to the chain axe from the oxygen side, it can be directly attached to the oxygen in the reaction C + $OCCH_2$ + M → $COCCH_2$ + M. In all these reactions, the water clusters are involved indirectly serving as a third body to which the reaction energy is transferred. Finally, this growth leads to the formation of rather large polymer molecules, which are not transferred into the gas phase after the evaporation of all helium, leading to a temperature rise. Inside dark molecular clouds, most of the hydrogen exists in a molecular form. Therefore, the reaction pathway, which involves dihydrogen instead of hydrogen atoms, should be rather efficient at high extinctions, and therefore it must be taken into account by modeling the reaction network leading to the formation of COMs by carbon atom addition.

As a result, in dense molecular clouds, low-temperature condensation of carbon would likely lead to the formation of COMs. Later, this organic material can be considerably modified by UV photon irradiation. The irradiation is expected to cause the release of CO, hydrocarbons, and small COM molecules into the gas phase. In the extreme case, almost all oxygen and hydrogen can be removed. Such processing should provide a rather amorphous carbon material. However, if the condensation of carbon takes place in diffuse molecular clouds, molecules physisorbed on the surface of the grains are efficiently desorbed by UV photons before they have a chance to react with next incoming atom or molecule. This is expected to lead to the formation of pure carbon grains, which might be followed by an instant temperature rise. As a result, the formation of a partially graphitized carbon grains is expected. Moreover, if the instant temperature rise is indeed triggered by collision of carbon clusters, this could lead to coalescence of carbon dust grains.

**Conclusions**

The low-temperature condensation of the vaporized graphite leads to the formation of carbon with fullerene-like structure and morphology similar to that produced by a high-temperature condensation. It is proposed that an instant temperature rise of carbon nanoparticles takes place during the condensation. Such a temperature rise could lead to the presence of a partially graphitized carbon grains in the dust formed in the diffuse ISM. The low-temperature condensation of carbon atoms together with the most abundant ISM molecules ($H_2$, $H_2O$, CO) leads to the formation of COM molecules, growth of which finally leads to the formation of refractory organic polymers. The absence of OH stretching vibrations in IR spectra of the condensate indicates that water molecules were not involved in the reaction network leading to the formation of COMs.

**Acknowledgments.** The authors are grateful for the support by the Austrian Science Fund FWF (P26635, W1259, M1908), Deutsche Forschungsgemeinschaft DFG (Contract No. KR 3995/3-1), and COST Action CM1401 "Our Astro-Chemical History". The authors acknowledge the help of A.V.Karabulin and V.I. Matyushenko in the experiments with the bulk liquid helium.

Appendix

Experimental and computational techniques used in the study

Figure 5 shows the experimental cell which was used in the study of the condensation of evaporated graphite. The cell was submerged into superfluid liquid helium kept at T = 1.7 K. The laser was focused on the place shown by the red star. The condensation of carbon took place mainly in the core of quantized vortices close to the ablation zone. The formed nanostructures were either pinned to the tops of the gilt electrodes or fallen down on the bottom of the cryostat, where the lacey carbon transmission electron microscopy (TEM) grids were mounted. In the cases of the metal or semiconductor coagulation, the sediment closed electrical circuit between the electrodes. However, no current was observed for carbon.

Figure 6 shows the principle scheme of the experiment on the condensation of atomic carbon together with $H_2O$, CO, and $H_2$ molecules. The helium droplets were generated by expansion of pressurized ($p$ = 20 bar) liquid helium through a 5 µm cooled nozzle. After the expansion, the evaporation of helium leads to the fragmentation of the jet and formation of helium nanodroplets. For the deposition on graphene the temperature of the nozzle was set to 6.5 K (that corresponds to the mean helium droplet sizes of about $10^8$ He atoms). The deposition time was 30 minutes. At these conditions the nanoparticles formed in helium droplets stay separate on a substrate. For the deposition on the fused silica substrates the temperature of the nozzle was set to 4.5 K and the deposition time was 180 minutes. Helium droplets containing few billions of He atoms were produced. Much larger helium droplet sizes resulted in a larger amount of material delivered to the substrate. Thus, coalescence of the nanoparticles on the silica substrates took place.

Quantum-chemical computations were performed to find whether the considered reactions are fast on the low-temperature surfaces of dust grains. The chemical composition as well as the morphology of the surface of the grains present in the ISM is not precisely known. Thus the catalytic activity of the surface cannot be taken into consideration correctly. Therefore, in our study, the computations were performed for gas-phase reactants. However, the main property of a surface (namely, the ability to absorb the reaction energy) should be considered. It allows the associative reactions A + B → AB. At the same time, it can result in freezing the reactions in intermediate local minima. Therefore, in order to test, whether a surface reaction is fast at low temperature, the presence of energy barriers and wells on the potential energy surface of reactions should be tested. For this purpose, first the geometries of the reactants were optimized. In the next step, the optimized structures were placed in the close vicinity to each other. A complete geometry optimizing procedure without any constrains was performed. The convergence of the system to the geometry of the final product was used as an indication of a fast reaction on cold inert surfaces.

| Element | Mass% | Atom % |
|---------|-------|--------|
| C       | 73    | 78     |
| N       | 0     | 0      |
| O       | 26    | 21     |
| Si      | ~ 1   | ~ 1    |
| Total   | 100.00 | 100.00 |

Table 1. EDX analysis of the deposit on the $SiO_2$ substrate.

Figure 1. TEM images of the carbon material produced by the condensation of vaporized graphite inside bulk superfluid liquid helium. The particles are free-standing over holes of the Lacey carbon film.

Figure 2. TEM images of the material condensed in helium droplets ($T_{nozzel}$ = 6 K, $P$ = 20 bar) and deposited on a graphene substrate.

Figure 3. IR absorption spectra of the material deposited on the fused silica substrate.

Figure 4. The initial chemistry pathway leading to the formation of COMs by addition of carbon atoms on interstellar grains proposed in previous studies and in this work.

Figure 5. The photo of submersible experimental cell used in the experiments on the condensation of vaporized graphite in superfluid bulk helium: 1 – graphite target; 2 – place of the laser focus; 3 – glass slide; 4 – the electrode array; 5 – TEM grids.

Figure 6. The principle scheme of the experiment on the condensation of atomic carbon together with common ISM molecules inside superfluid helium nanodroplets.

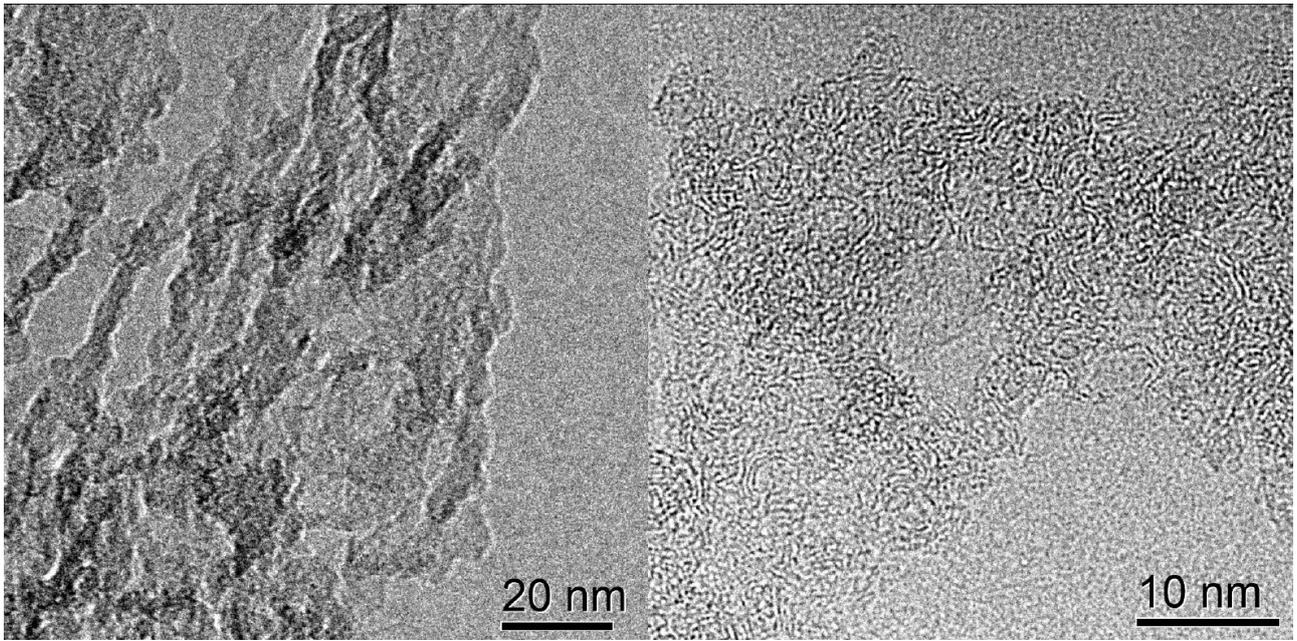

Figure 1

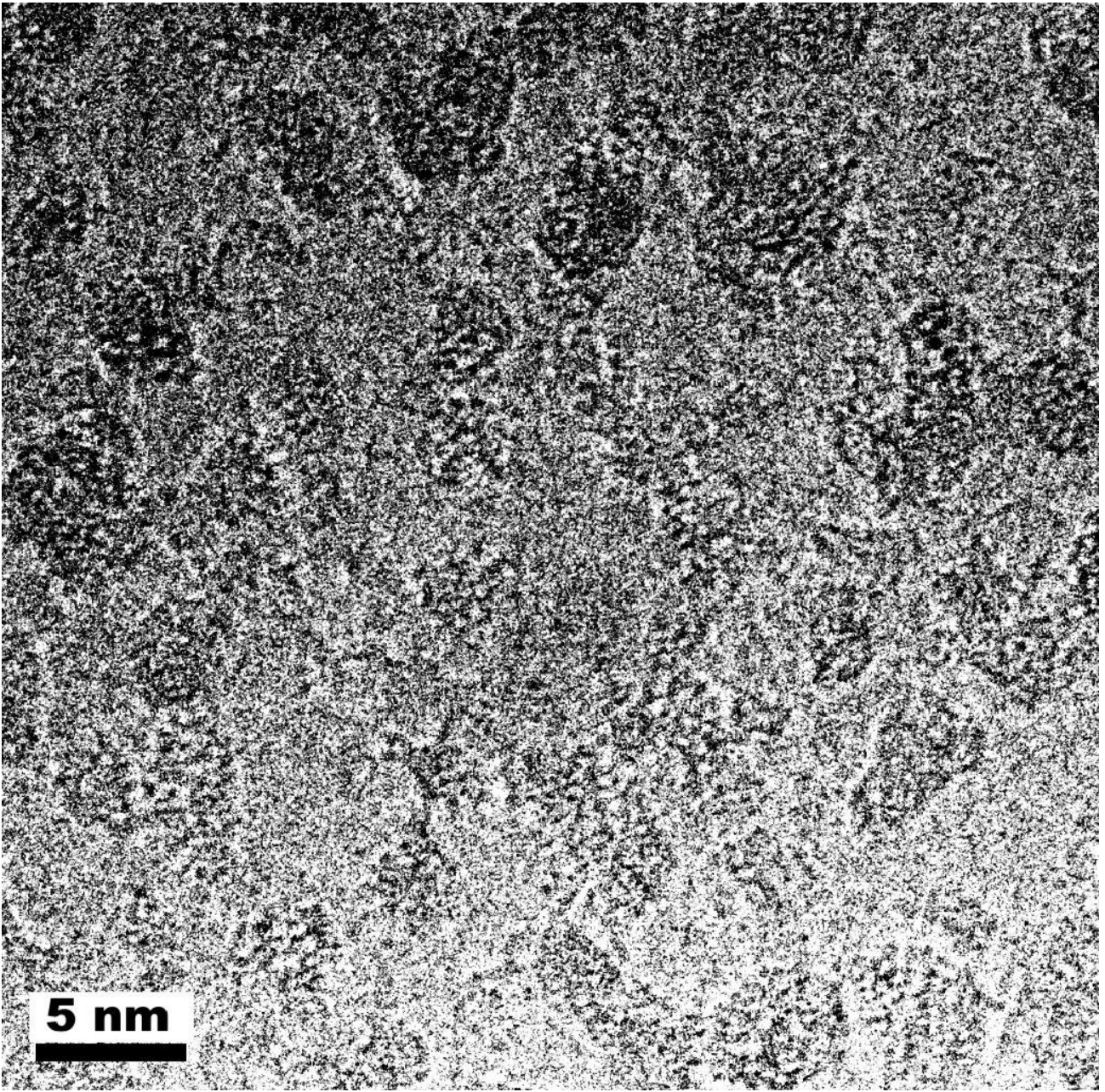

Figure 2

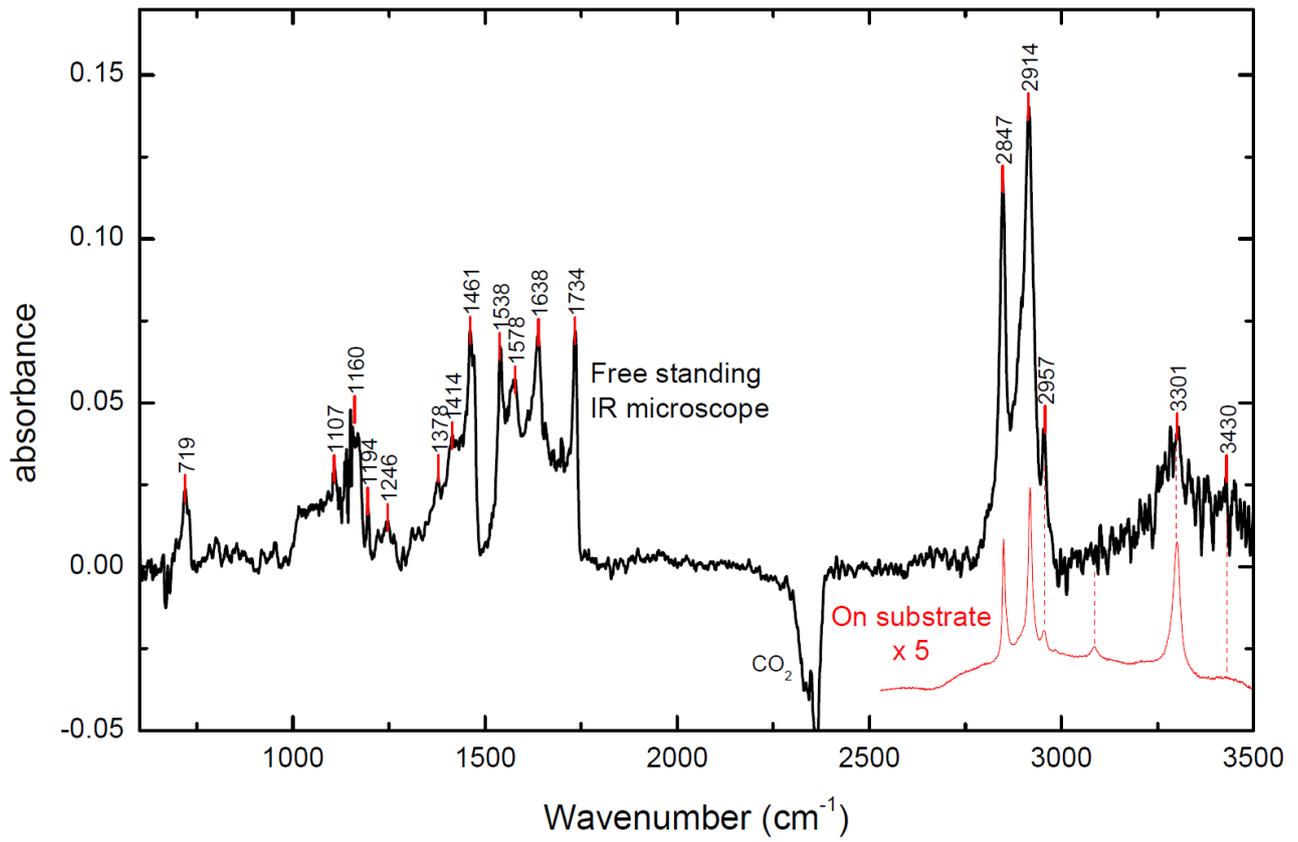

Figure 3

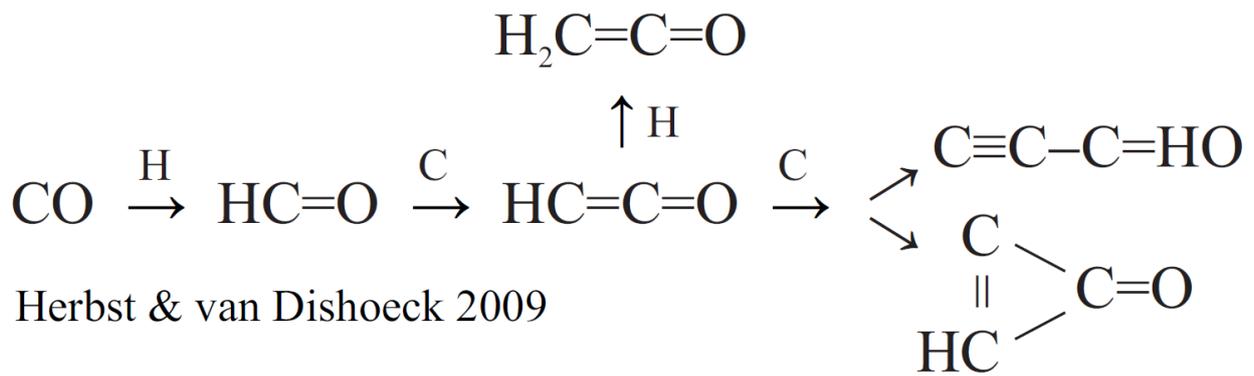

Herbst & van Dishoeck 2009

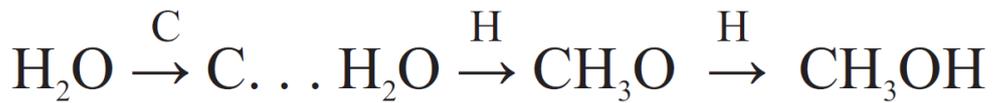

Ruaud et al. 2015

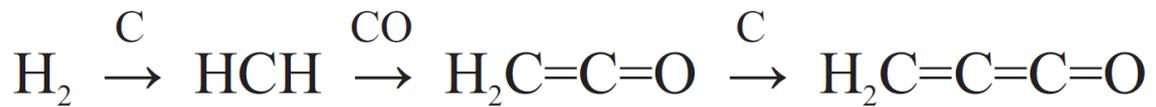

This work

Figure 4

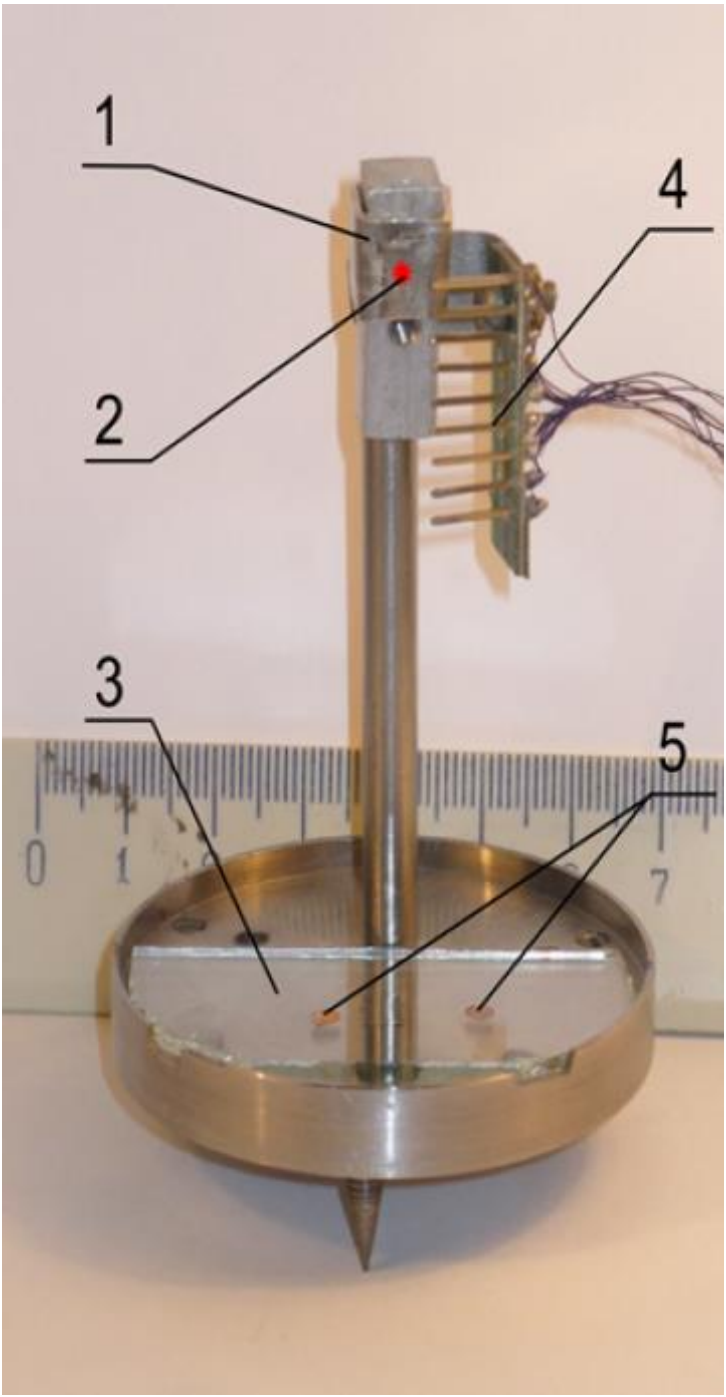

Figure 5

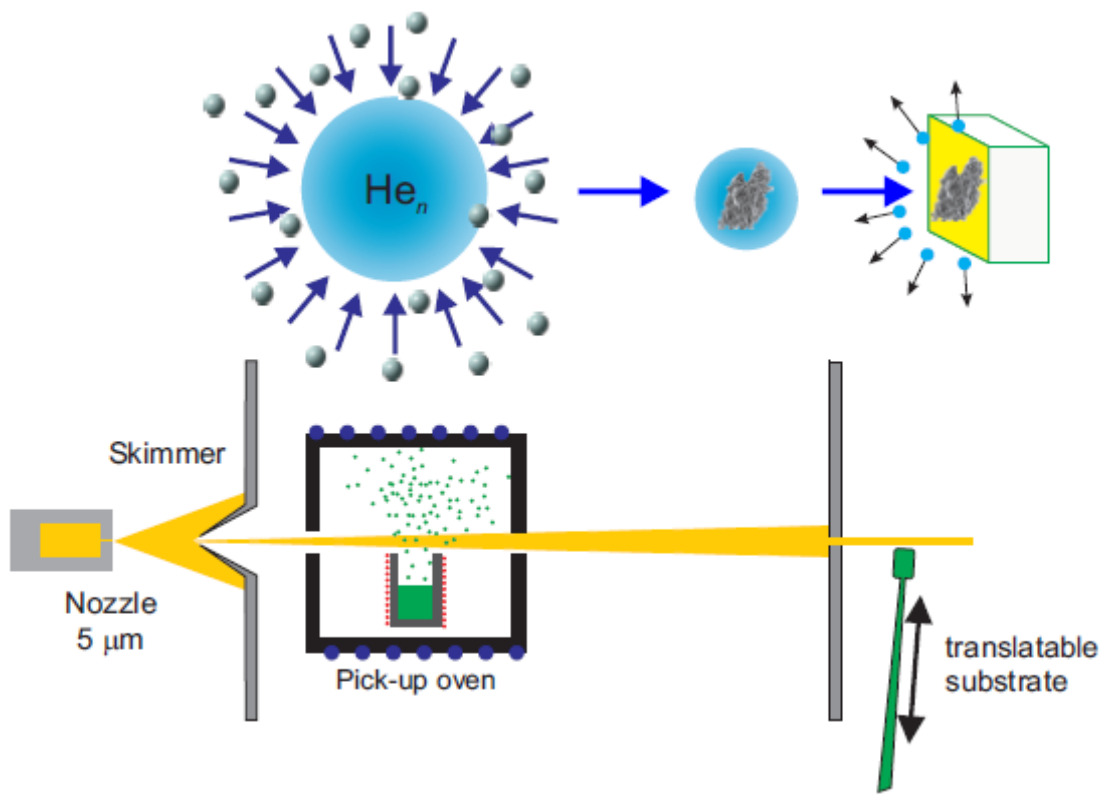

Figure 6